\documentclass[%
 reprint,
 nofootinbib,
 amsmath,amssymb,
 aps,
prl,
]{revtex4-2}

\pdfoutput=1

\usepackage{wrapfig}
\usepackage{blindtext}
\usepackage{multirow} 

\usepackage{graphicx}
\usepackage{tabularx}
\usepackage{rotating}
\usepackage{dcolumn}
\usepackage{bm}

\usepackage{amssymb}
\usepackage{pifont}
\usepackage{multirow}
\usepackage{mathrsfs}
\usepackage{mathtools}
\usepackage{makecell}
\usepackage{bbm}
\usepackage{amsmath}
\usepackage{bm}
\DeclarePairedDelimiter\bra{\langle}{\rvert}
\DeclarePairedDelimiter\ket{\lvert}{\rangle}
\DeclarePairedDelimiterX\braket[2]{\langle}{\rangle}{#1 \delimsize\vert #2}
\usepackage{mathtools}
\usepackage[toc,page]{appendix}

\usepackage{cellspace} %
\usepackage{siunitx} 

\usepackage[utf8]{inputenc}

\usepackage[margin=1in]{geometry}
\usepackage{datetime}
\usepackage{xcolor}

\newdateformat{monthdayyeardate}{%
  \monthname[\THEMONTH]~\THEDAY, \THEYEAR}

\usepackage{comment}

\begin{document}

\title{High-fidelity entanglement of metastable trapped-ion qubits with integrated erasure conversion}
\author{{A. Quinn}, {G. J. Gregory}, {I. D. Moore}, {S. Brudney}, {J. Metzner}, \\{E. R. Ritchie}, {J. O'Reilly}, {D. J. Wineland}, and {D. T. C. Allcock}}
\email[Corresponding author: ]{dallcock@uoregon.edu}
\affiliation{Department of Physics, University of Oregon, Eugene, OR, USA}

\date{\monthdayyeardate\today}
\begin{abstract}
Today's most advanced ion trap quantum computers have significant overhead due to the need for dual-species operation. Looking ahead, logical qubit register sizes will be limited by the encoding rate needed to correct generic Pauli errors. We address both of these issues by establishing high-fidelity control of metastable qubits, a key component of \textit{omg} or dual-type architectures, which enables converting a significant fraction of gate errors to erasures. We first implement an erasure conversion scheme which enables detection of $\sim 94\%$ of spontaneous Raman scattering errors during logic gates and nearly all errors from qubit decay.  Second, we perform a two-ion geometric phase gate using far-detuned (-44\,THz) stimulated Raman transitions to produce an entangled state with a raw Bell state fidelity of 97.73\% and a SPAM-corrected Bell state fidelity of 98.61\%.  When subtracting erasure errors, this fidelity becomes 99.16\%. These results, along with projections based on our detailed error budget, demonstrate metastable trapped-ion qubits as a platform for low-overhead, fault-tolerant quantum computing.
\end{abstract}

\maketitle

%
%
%
%

Trapped ions are a leading platform for quantum information processing (QIP), with the highest-recorded state preparation and measurement (SPAM)~\cite{An2022,sotirova2024high} and one- and two-qubit logic gate fidelities~\cite{Clark2021, löschnauer2024scalable,Smith2025}, but achieving true utility-scale quantum computing will likely require several orders-of-magnitude better logical performance~\cite{cao2019quantum}. This can, in principle, be achieved by leveraging fault-tolerant quantum error correction (QEC). The QEC overhead is significantly reduced when most errors are heralded leakage out of the computational subspace, also known as `erasure' errors~\cite{Wu2022,teoh2023dual,Kang2023}. A QEC code of a given size can generally correct twice as many erasures as Pauli errors and tends to have higher thresholds for erasures, thus providing a significant enhancement in logical fidelity versus code distance~\cite{grassl1997codes,stace2009thresholds,barrett2010fault,teoh2023dual}. Biases towards erasure errors have previously been engineered in both neutral atom~\cite{Ma2023} and dual-rail superconducting systems~\cite{levine2024demonstrating,chou2024superconducting,de2025mid}.

Typically, trapped-ion qubits are encoded within the electronic ground state manifold (\textit{g}) or across a narrow optical transition connecting the ground state to a metastable level (\textit{o}). Laser beams can be used to manipulate either of these encodings, but at finite power there is a fundamental error floor due to spontaneous photon scattering~\cite{Moore2023,Boguslowski2023} and, in the latter case, the metastable state lifetime. These events predominantly lead to undetected Pauli errors or leakage out of the qubit subspace that cannot be detected without significant overhead. Spontaneous Raman scattering (SRS) was the leading error source in the highest-fidelity demonstrations of stimulated Raman transition-based entangling gates for trapped ions~\cite{Gaebler2016,ballance2016}.

There has been recent interest in encoding qubits entirely within a metastable manifold (\textit{m}) where they are separated from the cycling optical transition radiation commonly used for laser cooling and fluorescence-based state readout~\cite{toyoda2009experimental,debry2023,Shi2025}. In the \textit{omg}~\cite{Allcock2021} or `dual-type'~\cite{Yang2022} architecture, this enables a separation of qubit functions (i.e. between sympathetic cooling and data storage) without the significant overhead of mixed-species ion crystals.
In addition, the spectral isolation between \textit{m} qubits and the ions used in the fluorescence cycle has been used to demonstrate record-high SPAM fidelities~\cite{sotirova2024high} and minutes-scale coherence times with simultaneous sympathetic cooling and continuous leakage monitoring~\cite{Shi2025}. In neutral atoms, metastable encodings have been shown to enable non-destructive mid-circuit measurement~\cite{Lis2023, Huie2023} and high-fidelity entangling gates where up to 33\% of errors were converted into erasures, limited by the complex branching structure out of the Rydberg states~\cite{Ma2023}.

The simple level structure of alkali-like ions with low-lying metastable $D$ states means that all leakage out of the metastable manifold ends up in $S_{1/2}$ or $D_{3/2}$. This population can be detected using standard fluorescence detection, at which point the error is considered an `erasure.' At the level of a single entangling gate, erasure conversion most naturally lends itself to post-selection, but erasure errors can also be corrected by standard quantum error correction codes with much greater efficiency than Pauli errors can be~\cite{Wu2022,Kang2023}. As the minimal example, at least four qubits are required to coherently correct a single erasure error as opposed to five for a Pauli error with unknown location~\cite{grassl1997codes}. \textit{m} qubits in the $D_{5/2}$ manifold of $^{40}$Ca$^{+}$ can be manipulated by driving coherent stimulated Raman transitions detuned from the $D_{5/2}\leftrightarrow P_{3/2}$ transition~\cite{Moore2023}. The branching ratio of the $P_{3/2}$ manifold ($\sim94\%$ to $S_{1/2}$~\cite{barton2000measurement}) dictates that most SRS errors can be converted into erasure errors as described above. Importantly, due to the spectral isolation of the \textit{m} qubits from the fluorescence cycle, this can be done with no impact on error-free qubits~\cite{Yang2022}. All errors caused by the finite metastable state lifetime can be detected in the same way~\cite{Shi2025}. This simple and convenient level structure makes high-rate erasure conversion much easier than in the Rydberg case~\cite{Ma2023}. A recent experiment demonstrated entanglement between two \textit{m} trapped-ion qubits with a Bell state fidelity of 96.3(6)\% and post-selection of erasure errors but did not quantify these errors or discuss limitations of the conversion process~\cite{wang2024experimental}.

In this Letter, we prepare an entangled state of two \textit{m} qubits in $^{40}$Ca$^{+}$ with a raw Bell state fidelity of 97.73(8)\% and a SPAM-corrected Bell state fidelity of 98.61(8)\%. By discarding 0.55(3)\% of the experimental shots, which correspond to erasure (i.e., detected) errors, we achieve a post-selected and SPAM-corrected Bell state fidelity of 99.16(7)\%. This corresponds to a 39\% reduction in the non-SPAM error via erasure conversion, which is competitive with the best demonstration in neutral atoms~\cite{Ma2023}. Furthermore, we provide a detailed characterization of both our erasure and non-erasure error sources and discuss experimental upgrades to mitigate both. Altogether, we elucidate the path towards \textit{m} qubits in trapped ions operating with a bias towards erasures, thus enabling more efficient QEC.

%
%

\begin{figure}[h]
\centering
    \hspace*{-0.0in}
    \includegraphics[scale = .52]{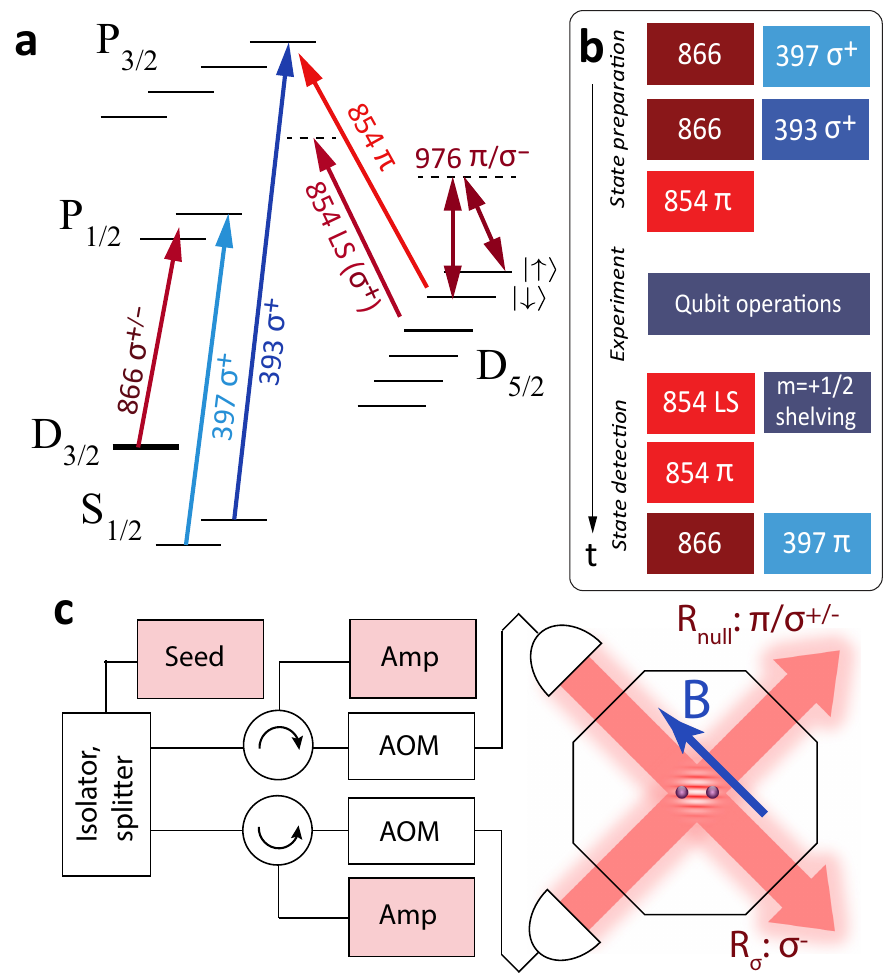}
    \caption{\textbf{a)} A level diagram for $^{40}$Ca$^{+}$ showing relevant transitions and their associated wavelengths (in nm).  \textbf{b)} Pulse sequence for a generic experiment in our \textit{m} qubit, showing the optical pumping technique used for state preparation and the shelving technique used for state detection. \textbf{c)} A schematic of our injection-locked diode laser system for producing 976\,nm beams for driving stimulated Raman transitions. Light from a 976\,nm free-space seed laser (0.7\,W) is injected into a pair of fiberized amplifier (amp) diodes (1.0\,W each) via optical circulators that in turn output into fiberized acousto-optic modulators (AOMs), which modulate the frequency of the fiberized light before output into free space optics at the trap (with powers up to 220\,mW per beam).}
\label{fig:experiment_setup}
\end{figure}

We perform gates on \textit{m} qubits encoded in the $\ket{\uparrow} \equiv \ket{m_J=+5/2}$ and $\ket{\downarrow} \equiv \ket{m_J=+3/2}$ levels of the $D_{5/2}$ manifold in $^{40}$Ca$^{+}$ (shown in Figure~\ref{fig:experiment_setup}a). The qubit splitting of 2.63\,MHz  is provided by the linear Zeeman effect ($|B|\approx$1.56\,G), with two-level dynamics ensured by using a $\sigma^+$-polarized laser beam tuned -21\,GHz from the $D_{5/2}\to P_{3/2}$ resonance to apply a light shift (LS) to states outside of the qubit manifold~\cite{sherman2013experimental}. This beam is labelled `854\,LS' in Figure~\ref{fig:experiment_setup} and later discussions. The qubits are prepared through optical pumping and read out using a state-selective deshelving~\cite{shelvingnote} process, the pulse sequences for which can be found in Figure~\ref{fig:experiment_setup}b.  Coherent qubit operations are performed with low scattering error rates using far-detuned (-44\,THz) 976\,nm stimulated Raman transition beams, with high powers supplied by a compact injection-locked diode laser system~\cite{Shimasaki2019} shown schematically in Figure~\ref{fig:experiment_setup}c and described in Section V of~\cite{supp_mat}.

%
%

\begin{figure*}
\centering
\includegraphics[width=\textwidth]{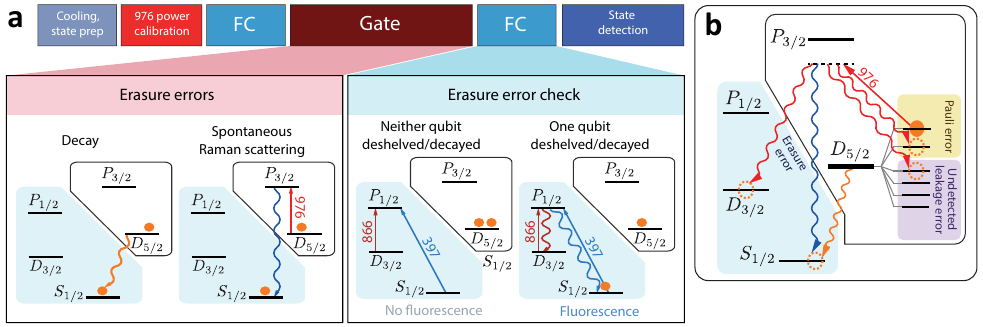}
\vspace{-7mm}
\caption{\textbf{a)} Schematic illustration of our scheme for converting leakage errors to erasures used with a two-ion gate, showing how FCs are used both before an algorithm to ensure that optical pumping has shelved both ions in $D_{5/2}$ and after for erasure conversion. \textbf{b)} Leakage pathways in our experiment, with SRS producing detectable leakage errors (labelled ``erasure errors"), undetected leakage errors, and Pauli errors.  Decay (in orange) due to the natural lifetime of the $D_{5/2}$ always produces a detectable leakage error.} 
\label{fig:leakage_detection}
\end{figure*}

Leakage out of the $D_{5/2}$ manifold can be detected by driving the $S_{1/2}\leftrightarrow P_{1/2}$ cycling transition and looking for 397\,nm fluorescence while preventing shelving in the $D_{3/2}$ manifold by driving the $D_{3/2}\rightarrow P_{1/2}$ transition, a technique we will refer to as a ``fluorescence check" (FC). Once leakage is detected, it is commonly referred to as an `erasure' error~\cite{grassl1997codes,Wu2022}. This process, in the context of a two-ion gate, is illustrated schematically in Figure~\ref{fig:leakage_detection}a. Leakage to other states in the $D_{5/2}$ manifold can in principle be detected by selectively deshelving these states using either narrow-linewidth $S_{1/2}\leftrightarrow D_{5/2}$ or 854\,nm $\sigma^+$-polarized beams prior to performing an FC, but as described below, such leakage errors should be relatively rare, and we do not perform this type of deshelving in the work reported here.

There are two primary leakage channels in our experiments, with the errors relevant to this experiment summarized in Figure~\ref{fig:leakage_detection}b: decay due to the finite natural lifetime $\tau=$1.16\,s~\cite{benhelm2008precision} of the $D_{5/2}$ manifold and SRS, which results in both leakage and spin flip errors, the relative probabilities of which are determined by the branching ratios out of $P_{3/2}$.  In our experiment, SRS deshelves the ion from the $D_{5/2}$ manifold in 94.1\% of scattering events, which can then be converted to erasure errors using a FC.  This is in contrast to \textit{g} qubits in $^{40}$Ca$^+$, for which scattering from the $P_{1/2}$ level will shelve the ion into the $D_{3/2}$ level in only $\sim$5\% of scattering events.  

\begin{figure}[h]
\centering
    \hspace*{-0.0in}
    \includegraphics[scale = .72]{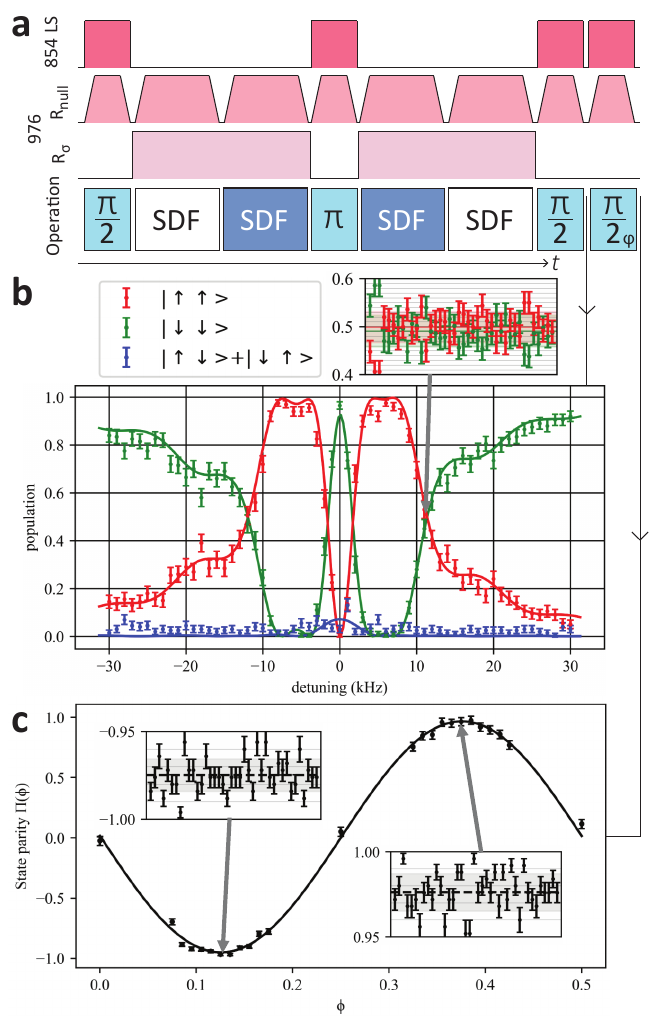}
    \caption{\textbf{a)} Pulse sequence for carrying out the gate, with SDF phase (0, $\pi$) denoted by color (white, blue). \textbf{b)} A plot of qubit state population versus SDF detuning from the motional mode resonance, showing the crossing point in $\ket{\downarrow\downarrow}$ and $\ket{\uparrow\uparrow}$ populations at which the gate was performed. \textbf{c)} A sample parity fringe.  Insets shows high-shot time series data at the gate operating point and parity fringe extrema.}
\label{fig:gate}
\end{figure}

We perform 1\,ms FCs after state preparation to improve SPAM fidelity (similar to the technique used in Ref.~\cite{sotirova2024high}) and prior to readout to check for leakage during the entangling gate.  Entanglement is mediated by the lower-frequency radial out-of-phase (``rocking") motional mode with resonant frequency $\omega_r =  2\pi\times 1.524$\,MHz which is first cooled to a Fock state occupation $\Bar{n}\approx0.1$ using electromagnetically induced transparency cooling~\cite{morigi2000ground} and pulsed sideband cooling~\cite{diedrich1989laser} on the \textit{m} qubits. 
The entangling geometric phase gate~\cite{Leibfried2003} pulse sequence is shown in Figure~\ref{fig:gate}a, and a representative plot of qubit state populations against detuning of the spin-dependent force (SDF) from the mode frequency resonance is shown in Figure~\ref{fig:gate}b. We drive a total of four phase space loops with a spin echo in the middle to perform Walsh modulation of both the motion (W(3)) and spin (W(1))~\cite{Hayes2012}. The total gate time is 400\,$\mu$s. The SDF is generated using a pair of orthogonal 976\,nm laser beams (see Figure~\ref{fig:experiment_setup}c).  Carrier operations are performed using a single bichromatic 976\,nm beam (labelled $R_\text{null}$) with a polarization that nulls the differential light shifts on the qubit levels. All $R_\text{null}$ pulses have a rise and fall time of 2\,$\mu$s to avoid off-resonant coupling between the spin and motion.

For the input state $\ket{\downarrow\downarrow}$, the ideal output of this gate is a Bell state $\ket{\Phi_+}=(\ket{\downarrow\downarrow}+\ket{\uparrow\uparrow})/\sqrt{2}$. The fidelity $\mathcal{F}=|\braket{\Phi_+}{\psi}|^2$ of the actual output $\ket{\psi}$ can be estimated by rotating it with a $\pi/2$ pulse performed at different phases $\phi$, producing a set of states with parity $\Pi=P(\downarrow\downarrow)+P(\uparrow\uparrow)-P(\uparrow\downarrow)-P(\downarrow\uparrow)$, ideally ranging from +1 to -1~\cite{Sackett2000}, where $P$ represents the population in each state~\cite{tomographynote}. A representative parity fringe is shown in Figure~\ref{fig:gate}c. To determine fidelity, we measure state populations at the end of the gate and the extrema of the parity fringes. At each of these points, we take 20,000 erasure-free shots, cycling between the three measurements every 500 shots. The associated time series are shown in insets on Figure~\ref{fig:gate}. From these measurements, we estimate the Bell state fidelity as $\mathcal{F}=(P_B(\downarrow\downarrow) + P_B(\uparrow\uparrow) + C)/2=0.9828(7)$ where $P_B$ are the populations and $C$ is the parity fringe contrast (half the difference between the fitted parity fringe extrema). Subtracting out a $87\times 10^{-4}$ two-ion SPAM error, this gives us a Bell state fidelity $\mathcal{F}=0.9916(7)$, with error bars calculated by bootstrapping the individual sets of 500 shot measurements~\cite{Clark2021}. We measure an additional erasure error rate of $55(3)\times10^{-4}$ that is excluded from the above figures because state measurement is aborted when these errors are detected. We summarize our estimated error contributions in Table~\ref{tab:errorbudget}.

\begin{table}[b]

\renewcommand{\arraystretch}{1.1}
\caption{\label{tab:budget}
Dominant sources of two-qubit gate infidelity as predicted by our error model. The lower half of the table summarizes the expected contributions due to erasure errors. Values in bold represent infidelity as measured in the gate. Aside from Raman scattering back into $D_{5/2}$, error sources contributing $<1\times 10^{-4}$ (including mode heating, carrier drive tuning errors, and uneven Raman beam intensities) are grouped together as ``Other contributions."}
\begin{ruledtabular}
\begin{tabular}{p{.05\textwidth}p{.25\textwidth}p{.08\textwidth}p{.05\textwidth}}

\textrm{} & \textrm{Error source}&
\multicolumn{2}{c}{\textrm{Infidelity $(\times10^{-4})$}}\\
\colrule
\multirow{7}{*}{\rotatebox[origin=c]{90}{Non-Erasure}}& Motional dephasing & 55\\
 & Spin dephasing & 26\\
 & $\pi$-time calibration & 6.8\\
 & Raman intensity drift & 4.8\\
 & Mode frequency drift & 1.9\\
 & Raman scattering to $D_{5/2}$ & 0.4\\
 & Other contributions $<10^{-4}$ & 2.6\\
\cline{2-4}
 & Total & 97.5$_{-0.09}^{+0.18}$ & \textbf{84(7)}\\
\colrule
\multirow{4}{*}{\rotatebox[origin=c]{90}{Erasure}} & $D_{5/2}$ lifetime (during gate) & 11.7\\
 & Raman scattering & 13.5\\
 & $854$ lightshift scatter & 5.6\\
 & Decay during FC & 17.1\\
 & Decay during prep FC & 6.8\\
\cline{2-4}
 & Total & 53.7(1) & \textbf{55(3)}
\label{tab:errorbudget}
\end{tabular}
\end{ruledtabular}
\end{table}

Of the expected non-erasure error we determine $87\times 10^{-4}$ arises from dephasing while nearly all of the remaining $11\times 10^{-4}$ is due to imperfect coherent processes (uncertainty in carrier $\pi$-time calibrations, beam intensity calibration errors, and mode frequency drift). See Section II of~\cite{supp_mat} for more details. The first class of error can be suppressed by improving coherence times or by running faster gates with increased Raman beam intensities or improved gate efficiencies, as discussed below.  Our dominant coherent errors, with the exception of mode frequency drift, are not directly reduced by increasing gate speed but can be addressed by other means (e.g. the use of composite pulses).

Erasure errors in our experiment stem from four factors.  Firstly, polarization impurities in the 854\,LS beam cause scattering out of the qubit states during carrier operations.  These errors are not fundamental to our gate and could be significantly reduced by improving polarization purity or increasing the 854\,LS detuning and power. More broadly, hyperfine clock qubits do not require a light shift for qubit isolation, meaning LS beams are not a general requirement for \textit{m} qubits.

Secondly, the 976\,nm beams induce SRS during carrier operations and spin-dependent force driving. We have characterized this error channel in a parallel publication~\cite{moore2025}. The gate time during which these events occur could be reduced by a factor of 17, thus achieving the $<10^{-4}$ error probability promised in Ref.~\cite{Moore2023}, but experimental design was constrained by practical considerations of flexibility and the need to mitigate dephasing. The relevant upgrades, discussed in more detail in Section III of~\cite{supp_mat}, are 1) encoding our qubit in the $m_J=+5/2$ and $m_J=-3/2$ states to maximize the difference in their Clebsch-Gordon coefficients, 2) forgoing Walsh modulation~\cite{Hayes2012} that protects against motional dephasing but increases the total gate time, 3) moving our Raman beams into a counter-propagating geometry, and 4) adding a second purely $\sigma^-$ Raman beam instead of relying on the $\sigma^-$ component of the $R_\text{null}$ beam. All of these upgrades would also significantly curtail the motional and spin dephasing error contributions.


Thirdly, our ion decays from the qubit manifold during gate operations, an error that can be reduced by reducing gate time, either by increasing gate efficiency (as discussed above) or by increasing Raman beam intensities.  Finally, decay during leakage checks imposes an error overhead unrelated to the gate itself.  Decay probability grows with the duration of the leakage check, introducing a trade-off between error overhead and the ability to accurately discriminate between dark and bright states as we discuss in Section IV of~\cite{supp_mat}. Decay during the pre-gate FC contributes to the observed erasure error but this is more properly considered a SPAM error.

To consider the error overhead produced by using an FC to detect leakage (as implemented in Figure~\ref{fig:leakage_detection}a), we calculate the breakdown of errors for a single ion with and without a final FC, accounting only for leakage and erasure errors.  If no FC is performed after running gates, the probability $P_l$ of experiencing an unheralded leakage error is $P_l = \epsilon_s + \epsilon_l$, where $\epsilon_s$ is the probability of the ion failing to shelve during state preparation and $\epsilon_l$ is the probability of deshelving during operations through decay or SRS.  When an FC is included, this leakage error will be partially converted to erasure error (which occurs with a probability $P_e$).  The efficiency of this conversion and contributions to $P_l$ and $P_e$ coming from the limited fidelity and finite lifetime of the FC are calculated in Section IV of~\cite{supp_mat}.

The predicted rates of these errors in our experiment (excluding $\epsilon_s$ since this amounts to a SPAM error) are presented in the first two columns of Table~\ref{tab:fc_error_table_main}.  The FC increases the total error $P_l+P_e$ by a factor of 1.6 while reducing unheralded leakage probability $P_l$ by a factor of 4.6.
Increasing photon collection efficiency and more tightly optimizing FC parameters could enable readout time to be reduced to $\sim$10\,$\mu$s~\cite{myerson2008high,crain2019high}, in which case we would expect the error overhead from decay during FC to be only $\sim 10^{-5}$.  In this context, added erasure error would be limited by the state detection fidelity, which can be as high as 99.9\%~\cite{myerson2008high,crain2019high}.

\begin{table}[b]
\renewcommand{\arraystretch}{1.1}
\caption{\label{tab:fc_error_table_main}
Single-ion error breakdown without FC, with FC as used in experiment, and with a hypothetical 10\,$\mu$s FC with a 99.9\% detection fidelity.  The value $P_l+P_e-\epsilon_l$ represents the change in total error from the no FC case.  Numbers are the error values multiplied by $10^4$.}
\begin{ruledtabular}
\begin{tabular}{p{.1\textwidth}p{.1\textwidth}p{.1\textwidth}p{.1\textwidth}}
Error & No FC & FC (expt.) & FC (10\,$\mu$s) \\
\colrule
Leakage, $P_l$ & 15.5 & 3.4 & $<$0.1\\
Erasure, $P_e$ & 0.0 & 20.7 & 15.6 \\
$P_l + P_e - \epsilon_l$ & 0.0 & 8.6 & 0.086 \\
\end{tabular}
\end{ruledtabular}
\end{table}

%
%
%
%

In summary, we have prepared \textit{m} qubits in $^{40}$Ca$^+$, used far-detuned stimulated Raman transitions driven by a compact laser system to entangle them with a record Bell state fidelity for metastable trapped ion qubits, and analyzed and demonstrated the use of fluorescence checks for highly-efficient erasure conversion. By quantifying all major error sources in our system, we are able to establish a road map to high-fidelity, erasure-biased operation of \textit{m} qubits with low-overhead mid-circuit sympathetic cooling and readout provided by the \textit{omg} architecture~\cite{Allcock2021}. By moving to metastable clock qubits in an isotope with non-zero nuclear spin~\cite{Shi2025} and improving the stability of our trap, we should enter a regime where erasure errors are the majority. We can further mitigate dephasing as well as metastable decay by adjusting our beam geometry or decreasing our Raman detuning. Finally, it is possible that optimal control could be used to convert non-erasure errors into erasures following the work in Ref.~\cite{Jandura2023}.

\textit{Acknowledgments} ---
We acknowledge useful discussions with C. Ballance, M. Boguslawski, W. Campbell, J. Chiaverini, I. Chuang, P. Drmota, T. Harty, E. Hudson, A. Hughes, A. Sotirova, and R. Srinivas as well as M. Polk's assistance on the apparatus.  This research is supported in part by the NSF through the Q-SEnSE Quantum Leap Challenge Institute, Award \#2016244 and the US Army Research Office under award W911NF-20-1-0037. The data supporting the figures in this article are available upon reasonable request from D.T.C.A.


\section{Supplemental Material}

\section{Gate Details}\label{appendix:gatedetails}

We perform a geometric phase gate in the $\sigma_z$ basis \cite{Leibfried2003}. As shown in Fig. 1 in the main text, we intersect two perpendicular $976$\,nm beams ($R_\sigma$ and $R_{\text{null}})$  with frequency difference $\omega$, generating a traveling interference pattern along the radial direction of the two-ion crystal.  Pulses of the $R_\text{null}$ beam are shaped, with a 2\,$\mu$s rise time, to minimize off-resonant driving of spin flips by the spin-dependent force (SDF) and off-resonant driving of the motion by the spin carrier. The travelling intensity pattern shifts the energy levels of the qubit via the AC Stark effect (i.e. a light shift), resulting in a spin-dependent force governed by the Hamiltonian
\begin{align}
    H=\frac{i\eta}{2}\sum_j(\Omega_{\uparrow}\ket{\uparrow_j}\bra{\uparrow_j} - \Omega_{\downarrow}\ket{\downarrow_j}\bra{\downarrow_j})e^{-i\delta t + i\phi}a^{\dagger} + \textrm{h.c.},
\label{gatehamiltonian}
\end{align}
\noindent where $j$ indexes the ion, $\eta$ is the Lamb-Dicke parameter, $\delta \equiv \omega - \omega_r$ is the detuning of the drive field from the lower frequency out-of-phase (OOP) mode, $\phi$ the phase of the drive field, and $\Omega_{\uparrow(\downarrow)}$ is proportional to the light shift on the $\uparrow(\downarrow)$ spin state. This spin-dependent force drives loops in motional phase space, accruing state-dependent geometric phases proportional to the loop area and generating entanglement.  To dynamically decouple slow noise and static calibration errors, we Walsh modulate the gate sequence~\cite{Hayes2012} in spin and motion. We carry out W(1) modulation on the spin with a spin-echo pulse in the middle of the gate sequence and a W(3) modulation on the motion, performing the gate in four phase space loops. A schematic of the full gate sequence is shown in Fig. 3a in the main text.  When characterizing gate output at different SDF detunings $\delta$, we modulate the phases between these loops such that the gate mode ideally returns to the motional ground state, disentangling the motion and the spin in a detuning-insensitive way. A representative plot of qubit state population over $\delta$ is shown in Figure 3b of the main text. 

\section{Error Budget}\label{appendix:errorbudget}
Here, we describe calculations used to estimate the dominant sources of two-qubit gate infidelity presented in Table I in the main text. We also detail sources of error not presented in the text, including mode heating, carrier drive tuning errors, and uneven
Raman beam intensities, all of which were found to be negligible compared with those in Table I in the main text.

\textit{Spin dephasing}. 
\newline
The Zeeman qubits used in this work suffer from spin decoherence due to their first-order sensitivity to ambient magnetic field fluctuations. To mitigate this, we implemented a Walsh 1 gate sequence in spin, reducing sensitivity to magnetic field noise that is slow on the timescale of the gate. Faster noise components however still degrade the coherence, and we found that loss of spin coherence accounted for a large portion of the total gate infidelity. To estimate the magnitude of this infidelity, we performed a Ramsey spin echo experiment on a single ion and measured the loss of Ramsey fringe contrast $c$ for a sequence of the same duration as the gate \cite{ballance2017high}. The corresponding two-qubit gate infidelity is $\frac{3}{2}\left(1-c\right)$. We measure a contrast loss $(1-c)$ of $17.5\times10^{-4}$ after subtracting single ion SPAM, corresponding to a gate infidelity of $26(2)\times10^{-4}$~\cite{ballance2017high}.
\newline

\begin{figure}
\centering
\includegraphics[scale=0.5]{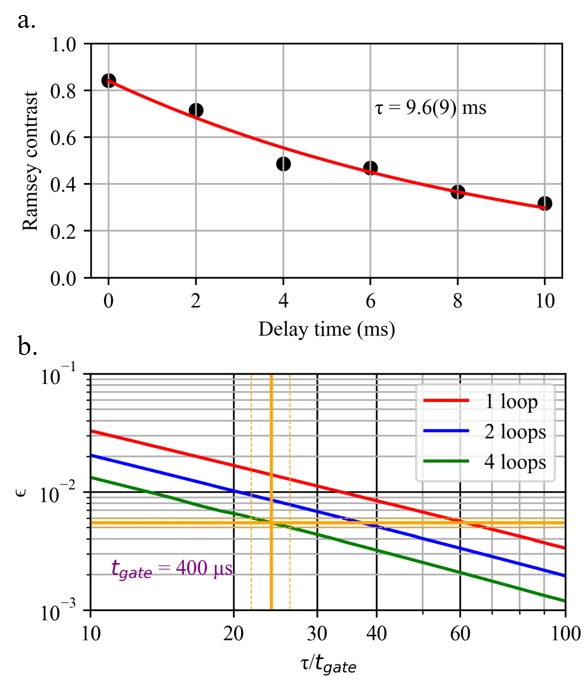}
\caption{\textbf{a)} Motional Ramsey experiment, plotting the contrast between the $\ket{0}$ and $\ket{1}$ fock states as a function of the delay time between sideband $\pi/2$ pulses. \textbf{b)} Numerically simulated Bell state infidelity due to a white motional noise source as a function of the coherence to gate time ratio. Orange lines represent the infidelity expected for measured coherence and gate times used in this work. Dashed lines represent the $68\%$ confidence interval. } 
\label{motionalramseypanel}
\end{figure}

\textit{Motional dephasing}
\newline
To estimate the Bell-state infidelity due to motional dephasing, we performed a motional analogue to the standard Ramsey experiment we utilized to characterize spin dephasing. By driving red and blue OOP sidebands, we first place the motional state of the crystal in a superposition of $\ket{0}$ and $\ket{1}$ Fock states. We then wait a delay time $T$ and perform the inverse of the motional $\pi/2$ pulse, scanning over the phase of that pulse to obtain a fringe contrast $c$ (see Figure~\ref{motionalramseypanel}a). We find an exponential decay (with coherence time $\tau$) well fits the Ramsey contrast as a function of the delay time $T$, suggesting approximately white motional frequency noise on the time/frequency scales relevant to the gate \cite{ballance2017high}. We then numerically simulate the Bell state infidelity as a function of ratio of the coherence time characterizing the white noise and the total gate duration by solving the master equations in Lindblad form numerically shown in Figure~\ref{motionalramseypanel}b. We fit a motional coherence time of 9.6(9) ms to the Ramsey contrast data, corresponding to a gate infidelity of $55(5)\times10^{-4}$.

\begin{figure}
\centering
\includegraphics[scale=.41]{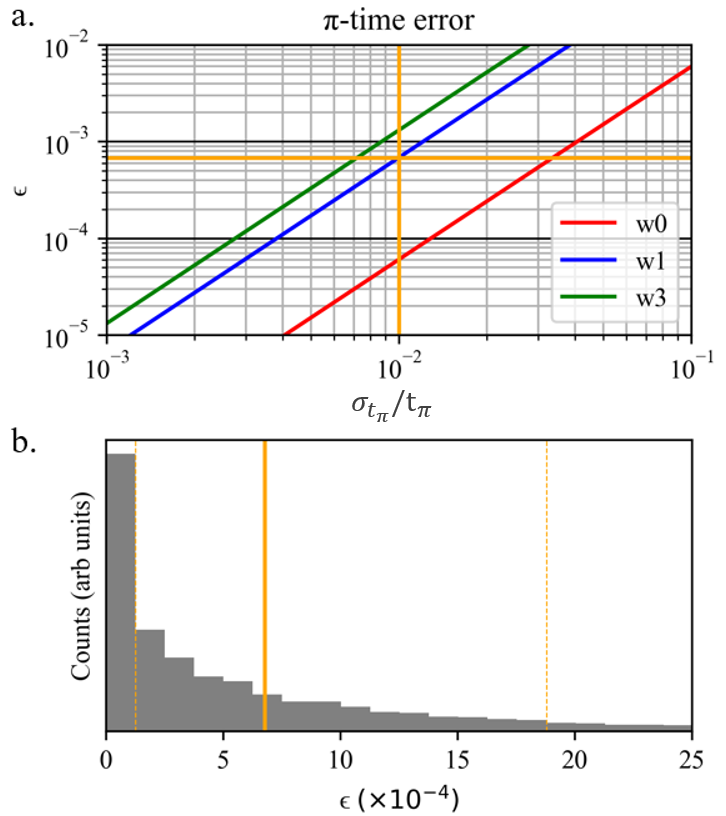}
\caption{\textbf{a)} Bell-state infidelity corresponding to a normally distributed relative $\pi$-time error characterized by standard deviation $\sigma_{t_\pi}$. \textbf{b)} Distribution of Bell-state infidelities corresponding to the measured $\sigma_{t_\pi}/t_\pi$ of $1\%$ in this work for a Walsh-1 gate. The solid orange line marks the average infidelity while the dashed orange lines represent the $68\%$ confidence interval.} 
\label{pitimepanel}
\end{figure}

\textit{$\pi$-time calibration}
\newline
Fits to determine the Raman-driven carrier $\pi$-times returned an error of $1\%$, suggesting that from shot to shot we can expect an error in the $\pi$-time calibration drawn from a normal distribution with a standard deviation of $1\%$. We plot numerical simulations of the Bell state infidelity due to percent offsets in the $\pi$-time drawn from a normal distribution characterized by $\sigma$, along with distributions of the infidelity to estimate the $68\%$ confidence interval (see Figure~\ref{pitimepanel}). Experimentally we found it difficult to control the relative phase of the carrier pulses between ions due to $\sim 1\%$ differences in Raman beam intensities at the ion that tended to drift on experimental timescales. To account for this in simulation we simply average over carrier phases. We find that for an error in the $\pi$-time of $1\%$ we expect a Bell-state infidelity of $6.8^{+12.0}_{-5.5}\times 10^{-4}$.

\textit{Raman intensity drift}
\newline
Alignment of the Raman beams generating the spin dependent force is subject to drifts over the course of the gate. Drifts of the beam along the axis of the crystal do not contribute any significant infidelity due to the symmetrized nature of the Walsh 1 sequence used in this work. However, drifts orthogonal to the axis of the crystal manifest as errors in the phase accrued to the $\ket{01}$ and $\ket{10}$ spin states. This can be seen as an asymmetry in the $\ket{00}$ and $\ket{11}$ populations after the gate sequence. To estimate the corresponding infidelity we numerically simulate the state populations as a function of error in Raman beam illumination, and infer an percentage illumination error of $1.3^{+1.0}_{-0.3}\%$ when comparing to the data. We then numerically simulate the Bell-state infidelity as a function of Raman illumination error, which we found to be $4.8^{+10.2}_{-2.1}$.

\begin{figure}
\centering
\includegraphics[scale=0.5]{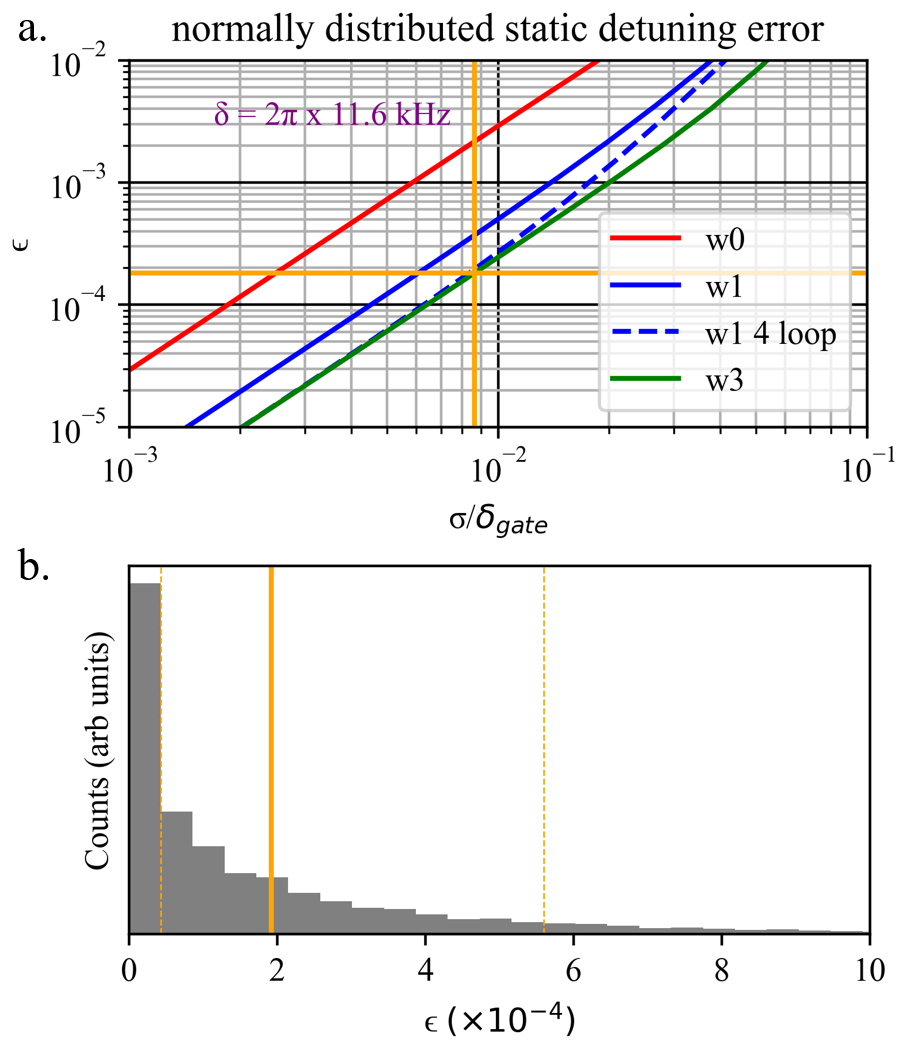}
\caption{\textbf{a)} Bell state infidelity corresponding to a normally distributed static offset in gate mode frequency with standard deviation $\sigma$ for different orders of Walsh modulation. \textbf{b)} Distribution of Bell state infidelities corresponding to a normal distribution of mode frequency offsets with a standard deviation of 100\,Hz for a Walsh 1 4-loop gate operating at $11.6$ kHz. Dashed orange lines represent $68\%$ confidence interval.} 
\label{modedriftpanel}
\end{figure}

\textit{Mode frequency drift}
\newline
The frequency of the mode on which the gate is driven is subject to slow drifts due to several technical reasons (e.g. thermal fluctuations in the trapping voltage source). To estimate the Bell state infidelity associated with this mode frequency drift, we first measured the width of the low frequency OOP mode to have a $\sigma$ of 100 Hz. We then numerically solved for the Bell state infidelity corresponding to an offset error in the gate mode frequency, and performed a weighted average over the infidelities corresponding to a normal distribution of offset errors characterized by a standard deviation $\sigma$ (see Figure~\ref{modedriftpanel}). To estimate the uncertainty in the infidelity, we looked at the distribution of infidelities generated by the distribution of mode frequency offsets with standard deviation $100$ Hz and simply located the $68\%$ confidence intervals of the infidelity distribution. From this method we found a Bell-state infidelity of $1.9^{+3.7}_{-1.4}\times 10^{-4}$.

\begin{figure}
\centering
\includegraphics[scale=0.63]{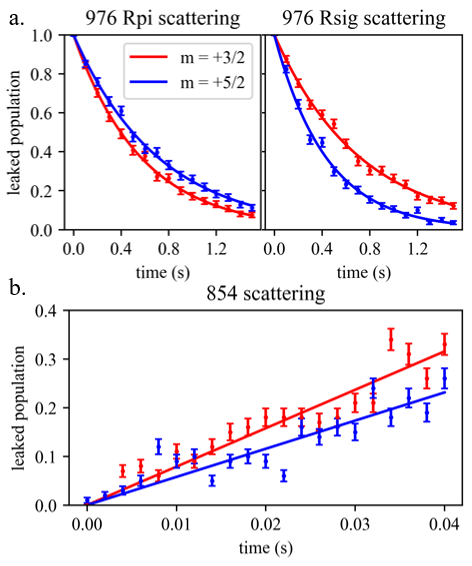}
\caption{\textbf{a)} Population scattered out of the $m=+5/2$ and $+3/2$  in the $D_{5/2}$ manifold with either the $976$ Rpi or Rsig beam on. \textbf{b)} Population scattered out of the qubit  states with the $854$ beam on.} 
\label{scatteringratepanel}
\end{figure}

\textit{Erasure errors}
\newline
To estimate infidelities from erasure error we simply measured the rate $\Gamma$ at which each beam scatters population out of the $D_{5/2}$ manifold for a single ion, as described in \cite{moore2025} and presented in Figure~\ref{scatteringratepanel}. Together with the duration over which the beam is on, the Bell state infidelity is $2\Gamma t$ (the factor of two accounting for two ions in the trap during the gate). Because both ions are in an equal spin state superposition throughout the gate, we measure the scattering rate out of each qubit state and consider the average scattering rate when calculating infidelity. 
We subtract the known natural lifetime decay rate of $1/1.16$\,Hz~\cite{benhelm2008precision} from each measured value before calculating the infidelity. We find Bell state infidelities of $13.5(2)\times10^{-4}$, $5.6(8)\times10^{-4}$, and $28.8(1)\times10^{-4}$ for erasure errors caused by 976 scattering, 854 scattering, and $D_{5/2}$ decay respectively. 

\textit{Raman/Rayleigh scattering to $D_{5/2}$}
\newline
We can then use the known branching ratios together with the measured scattering rates to calculate the scattering rates back into the $D_{5/2}$ manifold and using the same treatment to estimate the Bell state infidelity as was done for erasure errors. If however a qubit state scatters back to the same state (Rayleigh scattering) a recoil kick occurs and the Bell-state infidelity is instead $\frac{1}{2}\Gamma_{Rayleigh}t$~\cite{ballance2017high}. We find a Raman scattering error (non erasure) of $0.40(1)\times10^{-4}$ due to the $976$ beams, and infidelity $\ll10^{-5}$ for Rayleigh scattering errors.

\section{Reducing Raman Scattering Error}\label{appendix:ramanscattering}
Recent theoretical work \cite{Moore2023} predicts that, in calcium and with the -44\,THz Raman beam detuning used in our experiment, we should be able to run two-ion entangling gates with $<1\times 10^{-4}$ error rates from spontaneous Raman scattering.  In our gate however, we see a Raman scattering error of $(13.5(2)\times 10^{-4})$. Below, we explore the particulars of the experiment that raise this error rate and discuss changes that could be made to the beam geometry, beam polarizations, choice of qubit states in the $D_{5/2}$ manifold, and gate sequence that could bring Raman scattering error below $10^{-4}$.

\textit{Beam geometry}
Reduced efficiency from beam geometry comes from two factors.  Firstly, because the Raman beams are orthogonal, the wavevector difference $\Delta\Vec{k}$ between them (and the gate coupling rate) is a factor of $\sqrt{2}$ smaller than in the optimum case, where the beams are counterpropagating.  In addition, the motional mode used for the gate is at a 45$^\circ$ angle to $\Delta\Vec{k}$, reducing the coupling rate by a further factor of $\sqrt{2}$.  This means that changing the beam geometry could increase gate speed by up to a factor of two.

While the gate itself is driven by $\sigma^-$-polarized light, the beam $R_\text{null}$ contains other polarization components, added in order to enable carrier and sideband operations and to null the differential AC Stark shift from this beam on the qubit levels.  These components contribute to scattering but not to SDF strength, and because $\sigma^-$ makes up only one third of the total beam power in $R_\text{null}$, the other components reduce coupling rates by a factor of $\sqrt{3}$.  Nulling the AC Stark shift on the beam used to perform carrier operations eliminates the issue of carrier frequency miscalibration from beam drift and spin decoherence from beam power fluctuation, but is not necessary for performing the gate.  In addition, only a fraction of the total beam power (60\,mW out of 200\,mW) was used for driving carrier operations in the gate, and two thirds of this power was in polarization/frequency components that did not contribute to the drive.  In conjunction with the discussion about beam geometry above, a three-beam setup could be employed, with counter-propagating $\sigma^-$-polarized beams generating the SDF and a low-power $\pi$-polarized beam used for driving carrier operations and sidebands.

For the geometric phase gate performed, gate efficiency is directly related to the difference in the squared Clebsch-Gordon coefficients for the two qubit states. For $\sigma^-$ polarization and qubit states $m_J$ = +5/2 and $m_J$ = +3/2, this difference is 4/15, while if we encoded in $m_J$ = +5/2 and $m_J$ = -3/2 while keeping everything else the same, we would instead have a difference of 10/15. The ratio of these differences represents the speedup this switch would provide.  However, moving the lower-energy qubit state to a lower $m_J$ would both increase the magnetic field sensitivity and, for $m_J$=-3/2, require multi-step carrier pulses (e.g. $m_J$=+5/2$\to$+1/2 followed by +1/2$\to$-3/2).

To counteract spin and motional dephasing, a 4-loop, Walsh modulated gate was performed\cite{Hayes2012}.  For a $k$-loop gate, gate duration must increase by a factor of $\sqrt{k}$, so a 4-loop gate must be twice as long as a 1-loop one, which we could have performed with an otherwise faster gate and/or longer coherence times.

The avenues for gate speedup laid out above are not mutually exclusive, and were they all implemented, a factor of $\sim$17 reduction in gate time could be achieved, which would reduce the gate time from 400 to 23\,$\mu$s without changing the Raman scattering rate.

\begin{figure*}
\centering
\includegraphics[width=\textwidth]{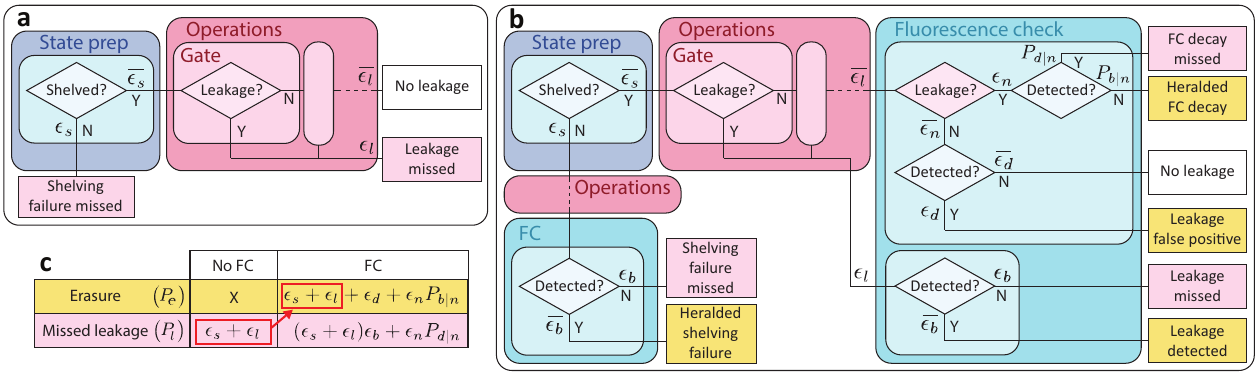}
\vspace{-7mm}
\caption{Flow charts illustrating the outcomes of a set of operations carried out on an \textit{m} qubit \textbf{a)} without a fluorescence check after gate operations and \textbf{b)} with a fluorescence check. Outcomes of individual events are marked with their associated probabilities.  Approximate total probabilities for erasures ($P_e$) and leakage errors ($P_l$) are shown in \textbf{(c)}. Outcomes classed as leakage errors are in pink, while erasures are in yellow.} 
\label{fig:FC_flowchart}
\end{figure*}

\section{Fluorescence Check Errors}\label{appendix:FCerrors}

Fluorescence checks (FC) for erasure conversion do two things: 1) convert leakage errors to erasure errors, and 2) introduce errors through decay of the metastable state (which can be detected or undetected) and through erasure false positives.  To quantify the contribution of this second effect to the total error, we can consider the possible states and detection outcomes for an experiment performed on a single ion with and without an FC prior to readout, as shown in Figure~\ref{fig:FC_flowchart}.  We can calculate these probabilities from a set of error probabilities: the probability $\epsilon_s$ of an ion not being shelved in the $D_{5/2}$ level by state preparation, the probability $\epsilon_l$ of a leakage event (decay or Raman scattering) deshelving the ion during gate operations, the probability $\epsilon_n$ of the ion decaying to the ground state during an FC due to the finite natural lifetime of the metastable manifold, and the readout error probabilities $\epsilon_d$ and $\epsilon_b$ for state detection of the dark and bright states, respectively.  Finally, $P_{d|n}$ and $P_{b|n}$ define the conditional probabilities of an FC identifying the ion as dark (undecayed) and bright (decayed) respectively given a decay event occurring during the FC.  

The outcome of an experimental run can be put into three categories: No error occurring, a leakage error occurring but not being detected (with probability $P_l$), and a leakage occurring and being converted into an erasure error (with probability $P_e$).  The size of these categories in different cases are discussed below.

\textit{Errors without FCs}
\newline
With no FC to look for leakage errors (shown in Figure~\ref{fig:FC_flowchart}a), $P_e=0$, and $P_l$ is just the sum of leakage errors occurring during gate operations 
\begin{align*}
    P_l &= \epsilon_s + \Bar{\epsilon}_s\epsilon_l \\
    &\approx \epsilon_s + \epsilon_l,
\end{align*}

\noindent where $\Bar{\epsilon}_s = 1 - \epsilon_s$.

\textit{Errors with FCs}
\newline
Introducing an FC to the end of gate operations increases the number of pathways leading to an error outcome (as shown in Figure~\ref{fig:FC_flowchart}b).  The probability of each outcome is simply the product of the probabilities of each branch leading up to that outcome, with three results corresponding to undetected leakages and four corresponding to erasure errors.  The total probabilities can be written
\begin{align*}
    P_l =& \Bar{\epsilon}_s \Bar{\epsilon}_l \epsilon_n P_{d|n} + \Bar{\epsilon}_s \epsilon_l \epsilon_b + \epsilon_s \epsilon_b, \\
    P_e =& \Bar{\epsilon}_s \Bar{\epsilon}_l \Bar{\epsilon}_n \epsilon_d + \Bar{\epsilon}_s \Bar{\epsilon}_l \epsilon_n P_{b|n} + \Bar{\epsilon}_s \epsilon_l \Bar{\epsilon}_b + \epsilon_s \Bar{\epsilon}_b.
\end{align*}

Keeping only the terms highest-order in these error contributions, these error probabilities can be approximated as
\begin{align*}
    P_l \approx & (\epsilon_s + \epsilon_l) \epsilon_b + \epsilon_n P_{d|n}, \\
    P_e \approx & \epsilon_s  + \epsilon_l + \epsilon_d + \epsilon_n P_{b|n}.
\end{align*}

Comparing the errors with and without the FC, it is clear that the effect of an FC is to take the $\epsilon_s + \epsilon_l$ leakage error from state preparation and qubit operations and to convert it to erasure error, with the conversion efficiency only limited by the readout error (specifically $\epsilon_b$).  The FC also introduces two additional types of errors: false positive erasure errors (represented by the $\epsilon_d$ term in the expression for $P_l$ above) and the error $\epsilon_n$ introduced by decay of the ion out of the metastable manifold during the FC (which, if caught by the FC, will be converted into an erasure error, but otherwise will be an unheralded leakage).  These errors can be non-negligible, with the second type discussed more thoroughly below.

\textit{Missed decays during FCs}
\newline
When $\epsilon_n$ is substantially smaller than $\epsilon_l$ or $\epsilon_s$ (e.g. if FCs are much shorter than the block of operations they follow, or if the time scales are comparable but SRS rather than decay is the leading source of leakage), then the inclusion of FCs will serve to convert unheralded leakage errors to erasures without contributing substantially to the total error. While this situation is possible, in our particular setup, the time taken to carry out FCs is on the same order of magnitude as the time taken to do the gate operations that we are interested in (roughly 1\,ms and 0.5\,ms respectively), and including errors from SRS, the probability of leakage during an FC is roughly half as large as the probability of leakage during the gate.  In this regime, the reduction in $P_l$ offered (e.g. whether or not the FC reduces unheralded leakage error by a factor of a few or by orders of magnitude) depends strongly on the probability $P_{b|n}$ of a decay event during an FC being detected during that same FC.

Given an FC with a length $t_0$, if a decay event happens at a time $t<t_0$, the associated expected photon count $\Bar{n}$ will be $\Bar{n}(t)=r_d t + r_b(t_0-t)$, where $r_d$ and $r_b$ are the photon count rates for the dark and bright states respectively.  Given that a decay event occurs at some point during the FC, the observed photon count distribution $\Bar{\Pi}(n)$ is just the sum of the Poisson distributions $\Pi(\Bar{n}, n)$ associated with each $\Bar{n}(t)$, weighted by the relative likelihood of decay over a time span $dt$ (which is uniform over the duration of the FC if $t_0 \ll \tau$), where $n$ is the observed photon number.  This can be written 
\begin{align*}
    \Bar{\Pi}(n) =& \frac{1}{t_0}\int_0 ^{t_0} dt \Pi(\Bar{n}(t), n), \\
    =& \frac{1}{\Delta\Bar{n}}\int_{\Bar{n}_d} ^{\Bar{n}_b} d\Bar{n} \Pi(\Bar{n}, n),
\end{align*}

\noindent where $\Delta \Bar{n} = \Bar{n}_b - \Bar{n}_d = r_b t_0 - r_d t_0$.  The probability $P_{b|n}$ of a correct identification of a decay event is then the probability of this distribution yielding a photon count higher than the threshold $n_t$, or
\begin{align*}
    P_{b|n} = \frac{1}{\Delta \Bar{n}}\sum_{n=n_t}^\infty  \int_{\Bar{n}_d} ^{\Bar{n}_b} d\Bar{n} \Pi(\Bar{n}, n).
\end{align*}

Aside from leakage errors introduced by the FC, an FC also introduces errors due to its inherently limited detection fidelity.  Undetected leakages should be rare so long as leakage error rates are low and bright state detection fidelity is high.  However, an error in detecting a dark state (i.e. measuring the ion as bright when it is actually dark) will identify an erasure error where none had occurred.  We class these errors are erasures because, if this misidentification is followed by a state reset procedure that deshelves the ion, then an actual erasure has occurred.

\textit{Prospects for optimization}
\newline
Checking for leakage will always increase the total error rate, suggesting that the optimal FC parameters (interrogation time and photon count threshold) are those which balance the cost of this increased error against the benefit of converting leakage to erasures.  Quantifying this rigorously is left to future work.  In our experiments, we selected parameters ($t_0$ and $n_t$ of 1\,ms and 25 respectively) that give a higher total error and a lower fractional reduction in $P_l$ than were possible but which make the FC more robust to fluctuations in associated beam parameters.

\begin{figure*}[h]
\centering
\includegraphics[width=\textwidth]{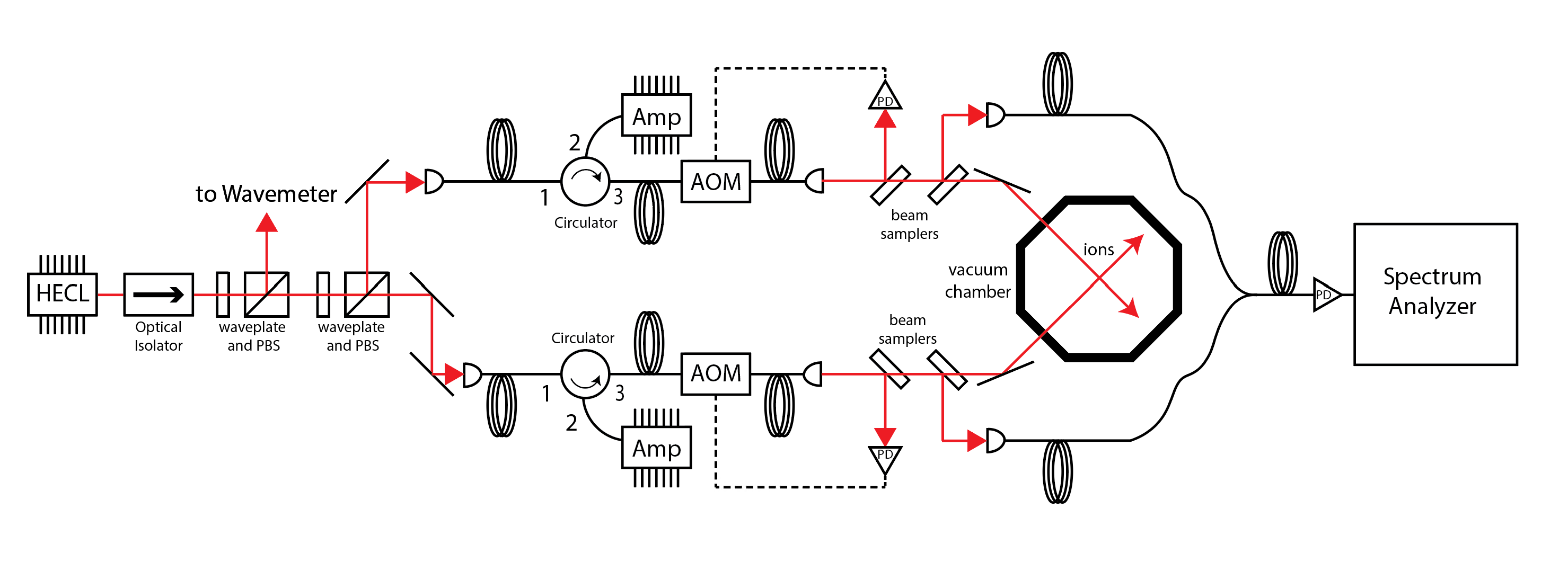}
\caption{Injection lock  optical setup. 
Solid red lines are free space laser beams with arrows for direction of travel.
Solid black lines and loops are fibers.
Dashed black lines represent the feedback system for maintaining consistent beam power. 
The second beam sampler in each beam path is used as a diagnostic tool for checking the quality of the injection lock.
This is removed during regular operation to maximize available power in each laser beam. 
} 
\label{fig:injection_lock_test_setup}
\end{figure*}

\section{Injection Locking}\label{appendix:injectionlocking}

We implemented an injection-locked diode system (Fig.~\ref{fig:injection_lock_test_setup}) based on~\cite{Shimasaki2019} in order to increase the optical power available in each of the 976\,nm laser beams. 
Compared to our previous system that used a single 700\,mW free space diode laser, this fiber-based injection-lock setup triples the optical power in each 976\,nm beam.
One major advantage is the cost-effectiveness of using small, butterfly-packaged high power diode lasers that were developed for the telecommunications industry as pump diodes for erbium-doped fiber amplifiers.

Light from a single VHG-based free-space 976\,nm 700\,mW hybrid external cavity laser (HECL) (the `seed' laser, 10976SB0700B from Innovative Photonic Solutions) passes through an optical isolator and a beamsplitter for diagnostics before it is split into two beam paths which are each coupled into Port 1 of two fiberized optical circulators (PMOC-F-3-98-5-PM980-90-05-FA from DK Photonics). 
About 50\,mW of light from the seed laser then exits each circulator from Port 2 and is injected into two fiber-pigtailed 976\,nm 1\,W diodes (the `amp' lasers, 976LD-2-0-0 from AeroDIODE).
The amp laser light re-enters the circulators from Port 2 and is directed out through Port 3 into fiberized acousto-optic modulators (AOMs) for switching and frequency tuning. The circulators prevent amplified light from exiting Port 1 and returning to the seed laser.

In free space, each beam is collimated with a beam waist diameter of $\sim$17\,mm by 60FC-L-4-M40-54 Schafter+Kirchhoff collimators. Beam samplers pick off 2.5\% of the light in each beam and direct it to photodiodes for intensity stabilization.
After the pickoffs, each beam is measured to have over 200\,mW, which is focused by lenses with focal length 150\,mm to $\sim$\SI{40}{\micro\metre} at the center of the ion trap.
The beatnote of two tones separated by 1\,MHz between the two AOMs is measured to have a full width at half maximum of 4\,Hz.  This test setup is shown in Figure~\ref{fig:injection_lock_test_setup}. 

\bibliography{refs}
\bibliographystyle{apsrev4-1}

\end{document}